\documentclass[letter,11pt]{article}
\usepackage[utf8]{inputenc}
\usepackage[margin=1in]{geometry}
\usepackage[]{graphicx}
\graphicspath{{figures/}}
\usepackage{caption, subcaption}
\usepackage{float}
\usepackage{color}
\usepackage{tikz}
\usepackage[normalem]{ulem}
\usetikzlibrary{automata,positioning,arrows}
\usepackage{amsmath}
\usetikzlibrary{calc}
\usepackage{bm}

\title{Heterogeneity of Interaction Strengths and Its\\
Consequences on Ecological Systems}
\author{Zachary Jackson, BingKan Xue}
\date{Department of Physics, University of Florida, Gainesville, FL, United States}

\begin{document}

\maketitle

\begin{abstract}

Ecosystems are formed by networks of species and their interactions. Traditional models of such interactions assume a constant interaction strength between a given pair of species. However, there is often significant trait variation among individual organisms even within the same species, causing heterogeneity in their interaction strengths with other species. The consequences of such heterogeneous interactions for the ecosystem have not been studied systematically. As a theoretical exploration, we analyze a simple ecosystem with trophic interactions between two predators and a shared prey, which would exhibit competitive exclusion in models with homogeneous interactions. We consider several scenarios where individuals of the prey species differentiate into subpopulations with different interaction strengths. We show that in all these cases, whether the heterogeneity is inherent, reversible, or adaptive, the ecosystem can stabilize at a new equilibrium where all three species coexist. Moreover, the prey population that has heterogeneous interactions with its predators reaches a higher density than it would without heterogeneity, and can even reach a higher density in the presence of two predators than with just one. Our results suggest that heterogeneity may be a naturally selected feature of ecological interactions that have important consequences for the stability and diversity of ecosystems.

\end{abstract}

\maketitle


\newpage
\section{Introduction} \label{sec:introduction}

Traditional physical systems involve interactions between objects characterized by universal coupling constants, such as the Newtonian constant of gravitation and the Coulomb constant for electrostatic forces. This is true whether the objects are subatomic particles, electric charges, or celestial bodies. The objects can differ in the amount of charge or mass they have, but otherwise interact in the same way with each other. This is in contrast to biological systems, where individual organisms are described by a large number of traits, be it morphological, metabolic, or behavioral, which are not all identical even between individuals of the same species \cite{bolnick:2011, siefert:2015, forsythe:2021}. Any of these traits can affect the way an individual interacts with other organisms and the environment. Thus, for biological interactions, the coupling constants (or ``interaction strengths'') themselves can be heterogeneous among individuals. Such interaction strengths can also vary in time due to behavioral changes, seasonal variations, or different life stages of an organism.

However, when trying to apply dynamical models to biological systems, the heterogeneity among individuals is often left out, so that a whole population is treated as having the same interaction strengths. A classic example is the Lotka-Volterra model of trophic interaction between two species, a predator and a prey. The predation rate is assumed to be proportional to the population densities of both species (in a quadratic form that loosely resembles some of the physical interactions mentioned above). This may be true if each population is homogeneous, so that only the population size or density is relevant. However, if we take into account the individual variation in various external or internal traits, the interaction strength between the predator and prey need not be the same for every individual. Intraspecific trait variation can have significant ecological effects \cite{bolnick:2003, bolnick:2011}. Thus, using an ``average'' interaction strength and ignoring the heterogeneity among individuals may cause models to miss important features. Here we address this problem by analyzing simple ecological models that demonstrate nontrivial consequences of heterogeneity in interaction strengths.  

As a case study, we analyze an ecosystem with ``exploitative competition'' between two predators feeding on the same prey species. Traditional models of such ecosystems treat each species as being homogeneous, so that the interaction strength between each pair is constant. We will incorporate heterogeneity in how the prey species interact with the predators, which can happen if there is trait variation among the prey population. Our model allows us to study several kinds of heterogeneity, including what we call ``inherent'', ``reversible'', and ``adaptive'' heterogeneities, depending on whether the prey phenotype is determined at birth, can change reversibly and stochastically, or adapts plastically to the density of predators. In all cases, we show that heterogeneous interactions lead to new phenomena in the population dynamics of species. Without heterogeneity, the expected outcome of the ecosystem is that one of the predators drives the other to extinction, a phenomenon known as ``competitive exclusion'' \cite{hardin:1960, chessen:2000, nguyen:2017}. However, if the interaction strength varies among the prey species, it turns out that the system can stabilize in a state where both predators coexist. For some range of parameters, we observe emergent facilitation between the predators \cite{deroos:2008}, such that the presence of one predator allows the other predator to persist. Moreover, the prey species can reach a higher abundance than when it interacts with only one predator, suggesting that heterogeneity in interactions may be an evolutionarily favorable feature for the population.

\section{System with homogeneous interactions} \label{sec:homogeneous}

\begin{figure}
\centering
\includegraphics[width=0.5\textwidth]{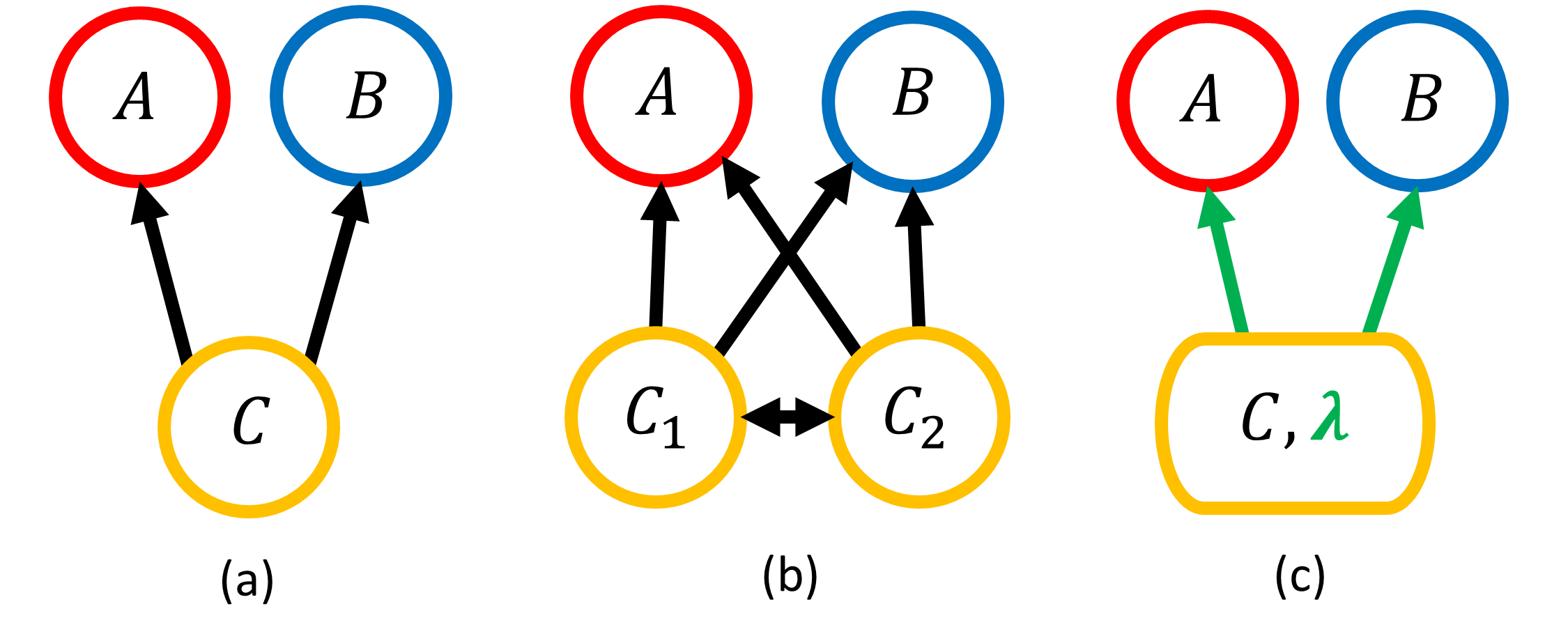}
\caption{\small An ecosystem with two predators ($A$, $B$) and one prey ($C$). (a) A homogeneous system where all individuals of a species have the same interaction strength with another species. Arrows represent trophic interaction pointing from prey to predator. (b) Heterogeneity is added by splitting the prey into two types ($C_1$, $C_2$), each with their own interaction strengths with the predators. Double-sided arrow represents the exchange of individuals between the prey subpopulations. (c) An equivalent description using the total prey population $C$ and the prey type composition $\lambda$. Arrows represent ``effective'' interaction strengths that depend on $\lambda$.}
\label{fig:variables}
\end{figure}

As a null model, we first consider an ecosystem with homogeneous interactions. The two predators and one prey species are modeled by a Lotka-Volterra dynamical system,
\begin{subequations}
\begin{align}
\dot{A} &= A \, (\alpha_{AC} \, C - \beta_A) \\
\dot{B} &= B \, (\alpha_{BC} \, C - \beta_B) \\
\dot{C} &= C \, (\beta_C - \alpha_{CC} \, C - \varepsilon_A \, \alpha_{AC} \, A - \varepsilon_B \, \alpha_{BC} \, B)
\end{align}
\end{subequations}
Here $A$, $B$, and $C$ represent the density of each species in a spatially well-mixed system, where $A$ and $B$ are the predators and $C$ is the prey (Fig.~\ref{fig:variables}a). Parameters $\beta_A$ and $\beta_B$ are the death rates of $A$ and $B$, respectively; $\alpha_{AC}$ and $\alpha_{BC}$ are the predation rates of $A$ and $B$ on $C$. Species $C$ has a birth rate $\beta_C$ and an intraspecific competition strength $\alpha_{CC}$. The $\varepsilon$'s are the efficiency by which an amount of $C$ consumed is converted to the increase of the predator population. We can rescale the variables by $t \leftarrow \beta_C \, t$, $A \leftarrow \frac{\varepsilon_A \alpha_{CC}}{\beta_C} A$, $B \leftarrow \frac{\varepsilon_B \alpha_{CC}}{\beta_C} B$, and $C \leftarrow \frac{\alpha_{CC}}{\beta_C} C$, so that the equations above simplify to
\begin{subequations}
\begin{align}
\dot{A} &= A \, (a \, C - a_0) \\
\dot{B} &= B \, (b \, C - b_0) \\
\dot{C} &= C \, (1-\, C - a \, A - b \, B)
\end{align}
\end{subequations}
where the rescaled parameters are $a = \frac{\alpha_{AC}}{\alpha_{CC}}$, $a_0 = \frac{\beta_{A}}{\beta_{C}}$, $b = \frac{\alpha_{AC}}{\alpha_{CC}}$, $b_0 = \frac{\beta_{B}}{\beta_{C}}$.

\begin{figure}
\centering
\includegraphics[width=0.5\textwidth]{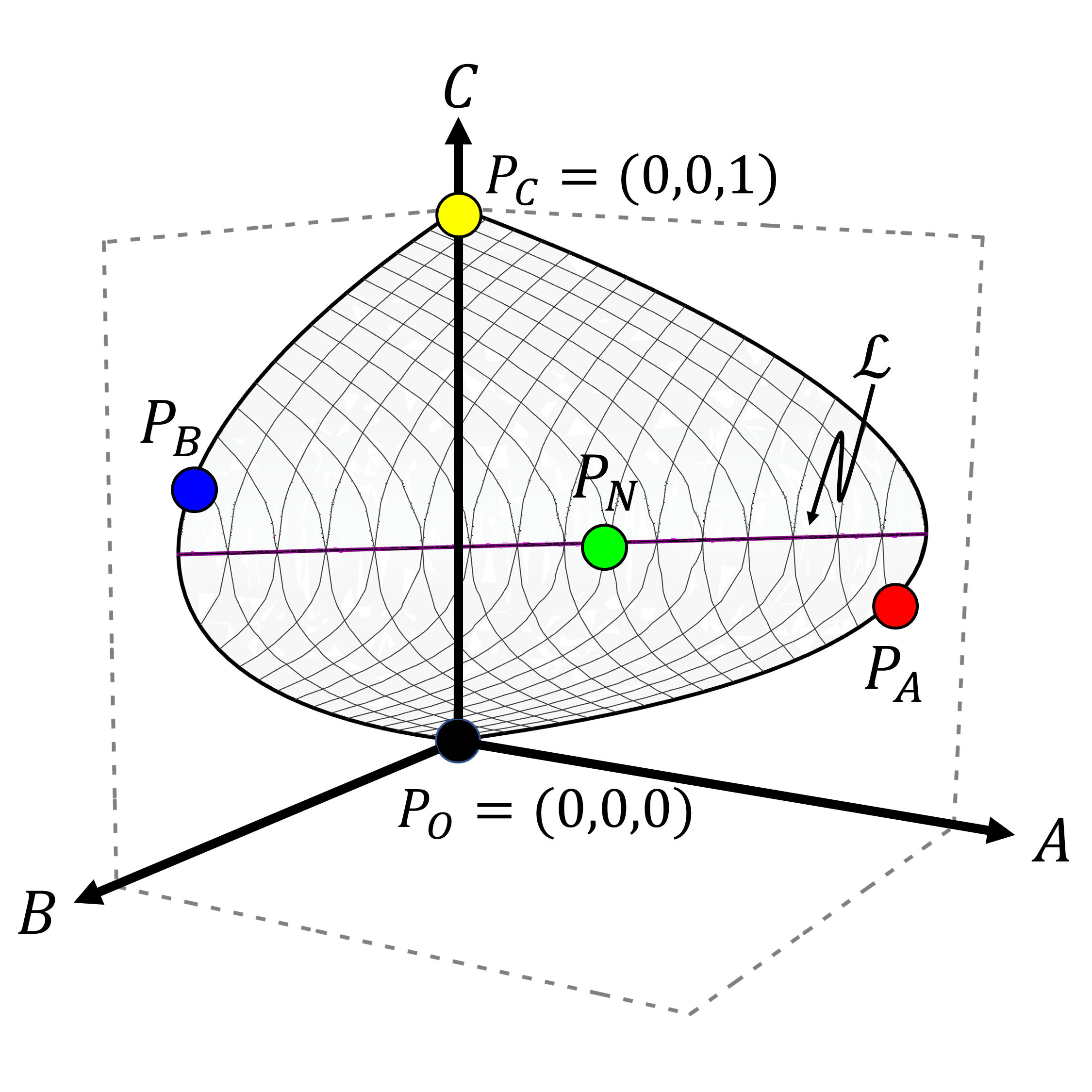}
\caption{\small $A$-$B$-$C$ space showing the locations of the equilibrium points. $P_O$ (black) is an unstable equilibrium at the origin where all species go extinct. $P_C$ (yellow) is where only $C$ survives. $P_A$ (red) and $P_B$ (blue) represent the persistence of one predator ($A$ and $B$, respectively) and the prey. These points each lie on a parabola in the $B=0$ and $A=0$ planes, respectively. $P_N$ (green) is a point where all three species coexist, which lies on a line $\mathcal{L}$ (purple) that is a line attractor for $\lambda = \lambda^*$. All five possible equilibria are on the surface $a_0 A + b_0 B = C (1-C)$, shown in gray.}
\label{fig:cbaphase}
\end{figure}

There are 4 equilibrium states of this system, which are labeled $P_O$, $P_C$, $P_A$, and $P_B$, as shown in Figure~\ref{fig:cbaphase}. They all belong to the surface $a_0 A + b_0 B = C (1-C)$ restricted to the non-negative octant of the $A$-$B$-$C$ space. $P_O$ is the point where all three species have zero population sizes, which is an unstable equilibrium. $P_C$ is where only $C$ persists; $P_A$ and $P_B$ are where $A$ or $B$ coexists with $C$, respectively. In general, only one of $P_A$, $P_B$, and $P_C$ can be stable for a given set of parameters. This demonstrates the competitive exclusion principle, by which two consumers (predators) of the same resource (prey) cannot coexist. The persistent predator is the one that has a lower $R^*$ value (i.e., the minimal prey density required to sustain a predator, $R^*_A = a_0 / a$ and $R^*_B = b_0 / b$). Only if the parameters are fine-tuned, such that the two predators have equal $R^*$, will all three species coexist. In such a fine-tuned system, there is a continuum of possible equilibria that form a line attractor $\mathcal{L}$ shown in Figure~\ref{fig:cbaphase}.

\section{Heterogeneous Interactions} \label{sec:heterogeneous}

Now consider a generalized model in which the species interactions are heterogeneous. A natural way of introducing heterogeneity in the system is by having a species diversify into subpopulations with different interaction strengths. This way of modeling heterogeneity is useful as it can describe different kinds of heterogeneity. For example, the subpopulations could represent polymorphic traits that are genetically determined or result from plastic response to heterogeneous environments. A population could also be divided into local subpopulations in different spatial patches, which can migrate between patches and may face different local predators. We can also model different behavioral modes as subpopulations that, for instance, spend more time foraging for food or hiding from predators. We will study several examples of these different kinds of heterogeneity after we introduce a common mathematical framework.

We will focus on the simple case where only the prey species splits into two types, $C_1$ and $C_2$, as illustrated in Figure~\ref{fig:variables}b. The situation is interesting when the predators $A$ and $B$ prefer different prey types such that, without loss of generality, $A$ consumes $C_1$ more readily while $B$ prefers $C_2$ (i.e., $a_1 / a_0 > b_1 / b_0$ and $b_2 / b_0 > a_2 / a_0$, which is equivalent to the condition that the nullclines of $A$ and $B$ cross, see Section~\ref{sec:nullcline}). The arrows between $C_1$ and $C_2$ in Figure~\ref{fig:variables}b represent the exchange of individuals between the two subpopulations, which can happen by various mechanisms considered below. Such exchange as well as intraspecific competition between $C_1$ and $C_2$ result from the fact that the two prey types remain a single species.

The system is now described by an enlarged Lotka-Volterra system with four variables, $A$, $B$, $C_1$, and $C_2$:
\begin{subequations}
\begin{align}
\dot{A} &= \alpha_{A1} A C_1 + \alpha_{A2} A C_2 - \beta_A A \\
\dot{B} &= \alpha_{B1} B C_1 + \alpha_{B2} B C_2 - \beta_B B \\
\dot{C_1} &= C_1 (\beta_C - \alpha_{CC} C)- \varepsilon_A \alpha_{A1} C_1 A- \varepsilon_B \alpha_{B1} C_1 B - \sigma_1 C_1 + \sigma_2 C_2 \\
\dot{C_2} &= C_2 (\beta_C - \alpha_{CC} C) - \varepsilon_A \alpha_{A2} C_2 A - \varepsilon_B \alpha_{B2} C_2 B + \sigma_1 C_1 - \sigma_2 C_2
\end{align}
\end{subequations}
Here we assume that the prey types $C_1$ and $C_2$ have the same intraspecific competition strength, but different interaction strengths with $A$ and $B$. The parameters in these equations and their meanings are listed in Table~\ref{tab:models}. For the convenience of analysis, we will transform the variables $C_1$ and $C_2$ to another pair of variables $C$ and $\lambda$, where $C \equiv C_1 + C_2$ is the total population of $C$ as before, and $\lambda \equiv C_2 / (C_1 + C_2)$ represents the composition of the population. After this transformation and rescaling of some variables (described in Appendix), the new dynamical system can be written as:
\begin{subequations}
\begin{align}
\dot{A} &= A \, \big( C (a_1 (1-\lambda) + a_2 \lambda) - a_0 \big) \label{eq:A-dot} \\
\dot{B} &= B \, \big( C (b_1 (1-\lambda) + b_2 \lambda) - b_0 \big) \label{eq:B-dot} \\
\dot{C} &= C \, \big( 1 - C - A (a_1 (1-\lambda) + a_2 \lambda) - B (b_1 (1-\lambda) + b_2 \lambda) \big) \label{eq:C-dot} \\
\dot{\lambda} &= \lambda (1-\lambda) \, \big( A (a_1 - a_2) + B (b_1 - b_2) \big) + \eta_1 (1-\lambda) - \eta_2 \lambda \label{eq:lambda-dot}
\end{align}
\end{subequations}
Here, $a_i$ and $b_i$ are the (rescaled) feeding rates of the predators on the prey type $C_i$; $a_0$ and $b_0$ are the death rates of the predators as before; $\eta_1$ and $\eta_2$ are the exchange rates of the prey types (Table~\ref{tab:models}). The latter can be functions of other variables, representing different kinds of heterogeneous interactions that we will study below.

\begin{table}
\centering
\begin{tabular}{||c |c| c||} 
\hline
Original &Rescaled & Meaning \\ [0.5ex] 
\hline\hline
$\alpha_{A1}$ & $a_1$ & consumption rate of $C_1$ by $A$ \\ 
\hline
$\alpha_{A2}$  & $a_2$ & consumption rate of $C_2$ by $A$ \\
\hline
$\alpha_{B1}$  & $b_1$ & consumption rate of $C_1$ by $B$   \\
\hline
$\alpha_{B2}$ & $b_2$ & consumption rate of $C_2$ by $B$  \\
\hline
$\alpha_{CC}$ & $1$ & intraspecific competition rate of $C$  \\
\hline
$\beta_A$ & $a_0$ & death rate of $A$  \\
\hline
$\beta_B$ & $b_0$ & death rate of $B$  \\
\hline
$\beta_C$ & $1$ & birth rate of $C$ \\
\hline
$\varepsilon_A$ & $1$ & biomass conversion of $C$ to $A$ \\
\hline
$\varepsilon_B$ & $1$ & biomass conversion of $C$ to $B$ \\
\hline
$\sigma_1$ & $\eta_1$ & exchange rate from $C_1$ to $C_2$ \\
\hline
$\sigma_2$ & $\eta_2$ & exchange rate from $C_2$ to $C_1$ \\
\hline
\end{tabular}
\caption{\small Parameters of the model (before and after rescaling) and their meanings.}
\label{tab:models}
\end{table}

\begin{figure}
\centering
\includegraphics[width=.5\textwidth]{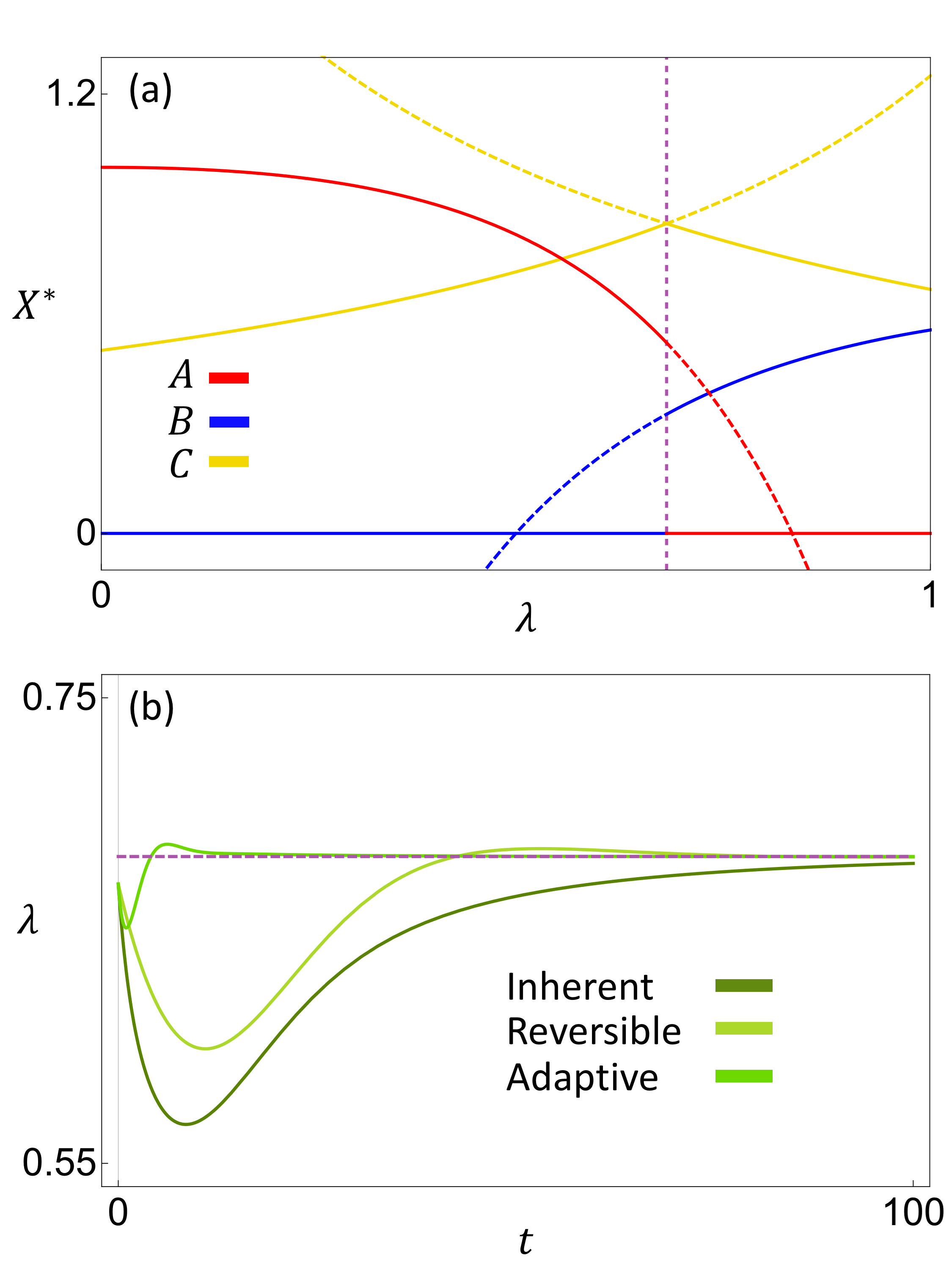}
\caption{\small (a) Equilibrium population of each species $X = A$, $B$, or $C$, with $\lambda$ held fixed at different values. Solid curves represent stable equilibria and dashed curves represent unstable equilibria. The vertical dashed line is where $\lambda = \lambda^*$, which is also the bifurcation point. Notice that the equilibrium population of $C$ is maximized at this point. (b) Time series of $\lambda$ for systems with each kind of heterogeneity. All three systems stabilize at the same $\lambda^*$ value, which is the bifurcation point in panel (a). Parameters used here are $(a_0,a_1,a_2,b_0,b_1,b_2,\rho,\eta_1,\eta_2,\kappa)=(0.25,0.5,0.2,0.4,0.2,0.6,0.5,0.05,0.05,50)$.}
\label{fig:lambda_dy}
\end{figure}

The variable $\lambda$ can be considered an internal degree of freedom within the $C$ population. In all models we study below, the dynamics of $\lambda$ is such that it stabilizes to a special value $\lambda^*$ (a bifurcation point), as shown in Figure~\ref{fig:lambda_dy}. As a result, a new equilibrium point $P_N$ appears in the $A$-$B$-$C$ space (on the line $\mathcal{L}$, see Figure~\ref{fig:cbaphase}), at which all three species coexist. If $\lambda$ were fixed at any other value, there would be exclusion of one of the predators (Figure~\ref{fig:lambda_dy}a), just as in the case of homogeneous interactions. Thus, with heterogeneous interactions, the prey composition $\lambda$ dynamically adjusts itself to the value $\lambda^*$ at which a new coexistence phase emerges, without having to fine-tune the interaction strengths. The exact conditions for this phase to be stable are detailed in the Appendix.

\subsection{Inherent heterogeneity} \label{sec:inherent}

We first consider a scenario where individuals of the prey species are born in two types with a fixed ratio, such that a fraction $\rho$ of the newborns are $C_2$ and $(1-\rho)$ are $C_1$. This could describe dimorphic traits, such as in sexual dimorphism \cite{frayer:1985,emlen:2005} where males and females are produced in a genetically determined ratio regardless of the composition of the population. We call this ``inherent'' heterogeneity, because individuals are born with a certain type and cannot change in later stages of life. The prey type given at birth will determine the individual's interaction strength with the predators. This kind of heterogeneity can be described by Eq.~(\ref{eq:lambda-dot}) with $\eta_1 = \rho (1-C)$ and $\eta_2 = (1-\rho) (1-C)$ (see Appendix).

\begin{figure}
\centering
\includegraphics[width=0.5\textwidth]{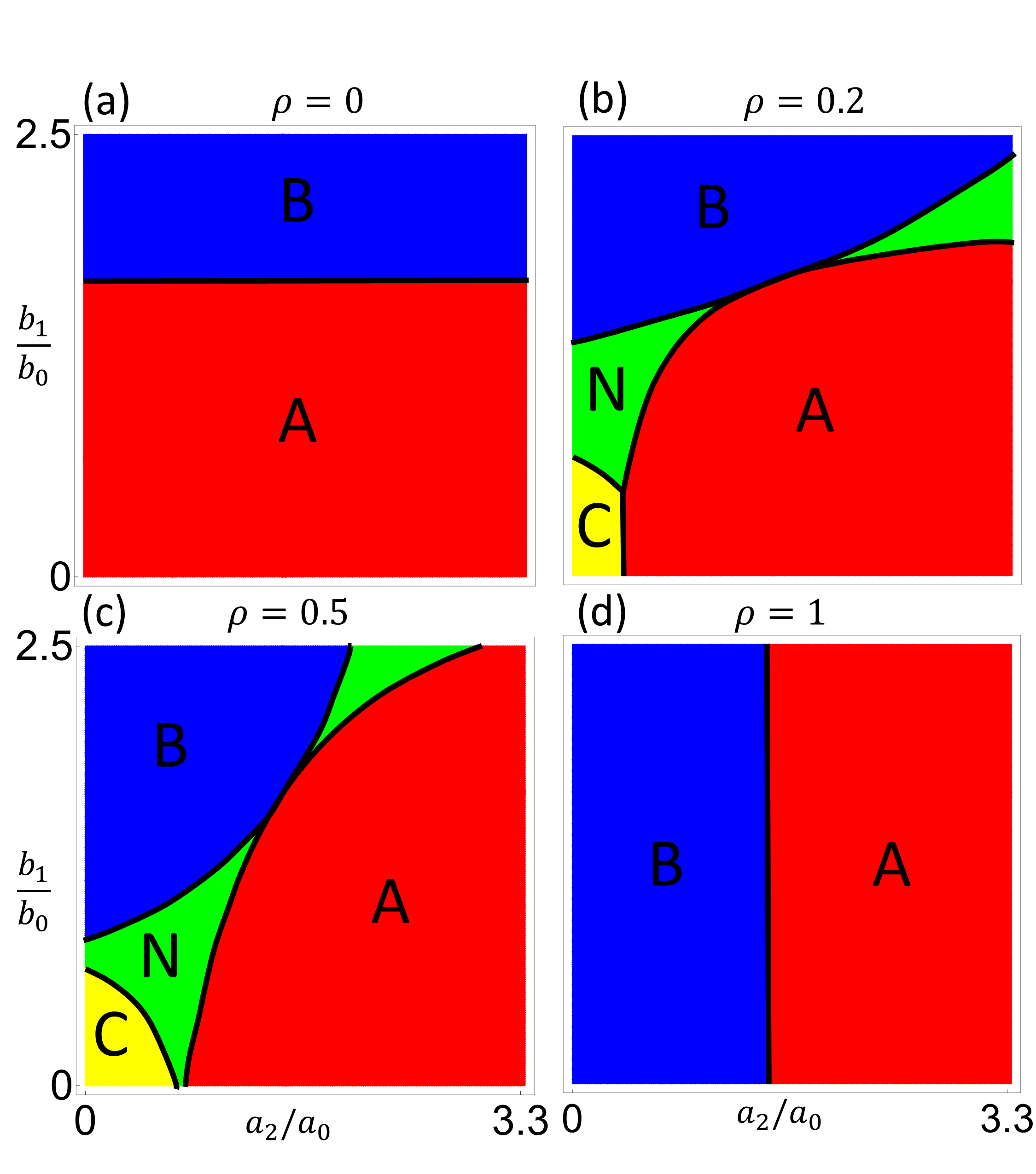}
\caption{\small Phase diagrams showing regions of parameter space identified by the stable equilibrium points. Yellow region represents $P_C$, red represents $P_A$, blue represents $P_B$, and green represents $P_N$. (a--d) Different values of the newborn composition $\rho$ for the model of inherent heterogeneity. In the extreme cases of $\rho = 0$ and $1$ the prey is homogeneous and there is no coexistence of the predators. Parameters used here are $(a_0, a_1, b_0, b_2) = (0.3, 0.5, 0.4, 0.6)$.}
\label{fig:diphase}
\end{figure}

The stable equilibrium of the system can be represented by phase diagrams that show the identities of the species at equilibrium. We plot these phase diagrams by varying the parameters $a_2$ and $b_1$ while keeping $a_1$ and $b_2$ constant. As shown in Figure~\ref{fig:diphase}, the equilibrium state depends on the parameter $\rho$. In the limit $\rho = 0$ or $1$, we recover the homogeneous case because only one type of $C$ is produced. The corresponding phase diagrams (Figure~\ref{fig:diphase}a,d) contain only two phases where one of the predators persist, illustrating the competitive exclusion principle. For intermediate values of $\rho$, however, there is a new phase of coexistence that separates the two exclusion phases.

\begin{figure}
\centering
\includegraphics[width=0.5\textwidth]{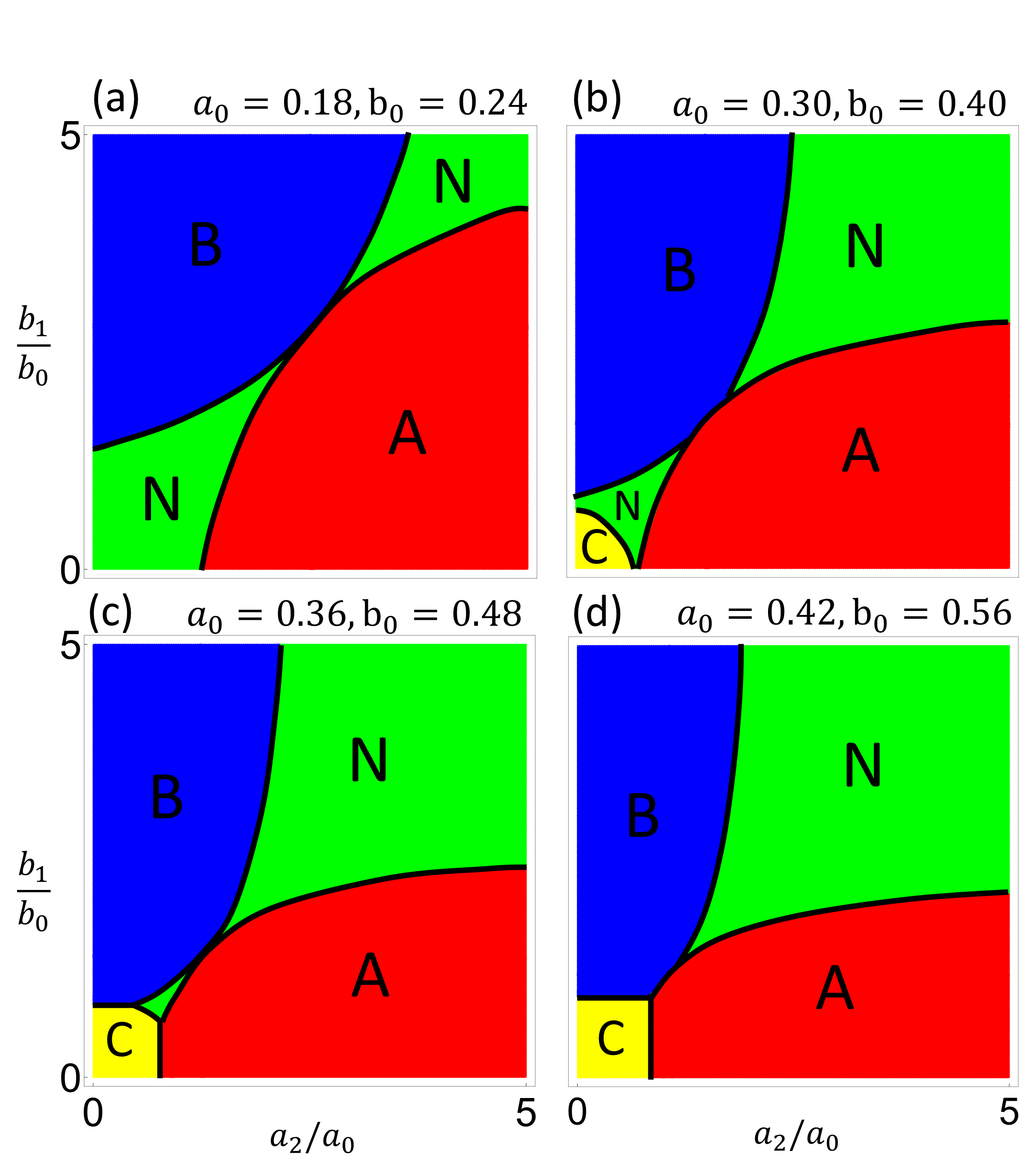}
\caption{\small Phase diagrams showing how the coexistence region changes with varying productivity (prey birth rate $\beta_C$). (a--d) The re-scaled predator death rates $a_0$ and $b_0$ are increased in proportion to each other, which is equivalent to decreasing $\beta_C$ (see Appendix for parameter transformation). The lower left region of coexistence (green) is absorbed into the prey-only region (yellow) as productivity is decreased. Parameters used here are $(a_1, b_2, \rho) = (0.5, 0.6, 0.5)$.}
\label{fig:prod}
\end{figure}

There are two such regions of coexistence, which touch at a middle point and open toward the bottom left and upper right, respectively. The middle point is at $(a_2/a_0 = b_2/b_0, b_1/b_0 = a_1/a_0)$, where the feeding preferences of the two predators are identical (hence their niches fully overlap). Towards the origin and the far upper right, the predators consume one type of $C$ each (hence their niches separate). The coexistence region in the bottom left is where the feeding rates of the predators are the lowest overall. There can be a region (yellow) where both predators go extinct, if their primary prey type alone is not enough to sustain each predator. As shown in Figure~\ref{fig:prod}, increasing the productivity of the system by increasing the birth rate ($\beta_C$) of the prey eliminates this extinction region, whereas lowering productivity allows the extinction region to subsume the lower coexistence region.

The new equilibrium is not only where the predators $A$ and $B$ can coexist, but also where the prey species $C$ grows to a larger density than what is possible for a homogeneous population. This is illustrated in Figure~\ref{fig:lambda_dy}b, which shows the equilibrium population of $C$ if we hold $\lambda$ fixed at different values. The point $\lambda = \lambda^*$ is where the system with a dynamic $\lambda$ is stable, and also where the population of $C$ is maximized. That means the population automatically stabilizes at the optimal composition of prey types. Moreover, the value of $C^*$ at this coexistence point can even be larger than the equilibrium population of $C$ when there is only one predator $A$ or $B$. This will be discussed further in Section~\ref{sec:promotion}. These results suggest that heterogeneity in interaction strengths can potentially be a strategy for the prey population to leverage the effects of multiple predators against each other to improve survival.

\subsection{Reversible heterogeneity} \label{sec:reversible}

We next consider a scenario where individuals can switch their types. This kind of heterogeneity can model reversible changes of phenotypes, i.e., trait changes that affect the prey's interaction with predators but are not permanent. For example, changes in coat color or camouflage \cite{martin:2009, niu:2018, noor:2008}, physiological changes such as defense \cite{mickalide:2019}, and biomass allocation among tissues \cite{minichin:1994, gedroc:1997}. One could also think of the prey types as subpopulations within different spatial patches, if each predator hunts at a preferred patch and the prey migrate between the patches \cite{nicholson:1935, hastings:1994}. We can model this ``reversible'' kind of heterogeneity by introducing switching rates from one prey type to the other. In Eq.~(\ref{eq:lambda-dot}), $\eta_1$ and $\eta_2$ now represent the switching rates per capita from $C_1$ to $C_2$ and from $C_2$ to $C_1$, respectively. Here we study the simplest case where both rates are fixed.

\begin{figure}
\centering
\includegraphics[width=0.5\textwidth]{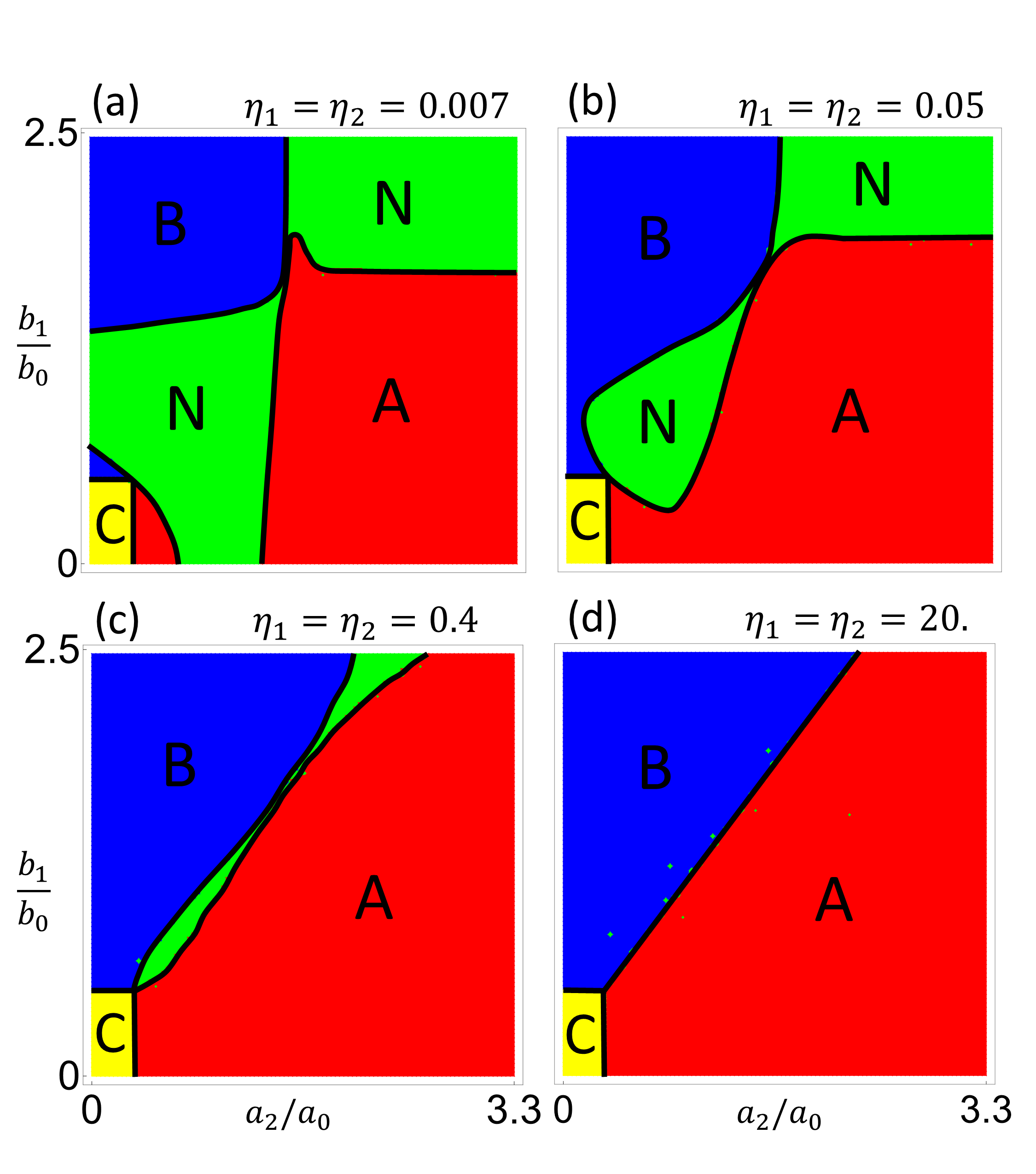}
\caption{\small Phase diagram showing how the coexistence region changes with the switching rates $\eta_1$ and $\eta_2$. (a--d) The coexistence region shrinks as the switching rates increase. When these rates are large, $\lambda$ is effectively held constant, so the system is equivalent to a homogeneous system and coexistence disappears. Parameters used here are $(a_0, a_1, b_0, b_2) = (0.3, 0.5, 0.4, 0.6)$}
\label{fig:phenophase}
\end{figure}

In the absence of the predators, the natural composition of the prey species given by the switching rates would be $\rho \equiv \eta_1 / (\eta_1 + \eta_2)$, and the rate at which $\lambda$ relaxes to this natural composition is $\gamma \equiv \eta_1 + \eta_2$. Compared to the previous scenario where we had only one parameter $\rho$, here we have an additional parameter $\gamma$ that modifies the behavior of the system. Figure~\ref{fig:phenophase} shows phase diagrams for the system as $\rho$ is fixed and $\gamma$ varies. We again find the new equilibrium $P_N$ where all three species coexist. When $\gamma$ is small, the system has a large region of coexistence. As $\gamma$ is increased, this region is squeezed into a border between the two regions of exclusion. However, this is different from the exclusion we see in the limit $\rho \to 0$ or $1$ for the inherent heterogeneity, where the borders are horizontal or vertical (Figure~\ref{fig:diphase}). Here the predators exclude each other despite having a mixture of prey types in the population.

This special limit can be understood as follows. For a large $\gamma$, $\lambda$ is effectively set to a constant value equal to $\rho$, because it has a very fast relaxation rate. In other words, the prey types exchange so often that the population always maintains a constant composition. In this limit, the system behaves as if it were a \emph{homogeneous} system with effective interaction strengths $a_\text{eff} = (1-\rho) \, a_1 + \rho \, a_2$ and $b_\text{eff} = (1-\rho) \, b_1 + \rho \, b_2$. As in a homogeneous system, there will be competitive exclusion between the predators instead of coexistence. This demonstrates that having a constant level of heterogeneity is not sufficient to cause coexistence. The overall composition of the population must be able to change dynamically as a result of population growth and consumption by predators.

\begin{figure}
\centering
\includegraphics[width=0.5\textwidth]{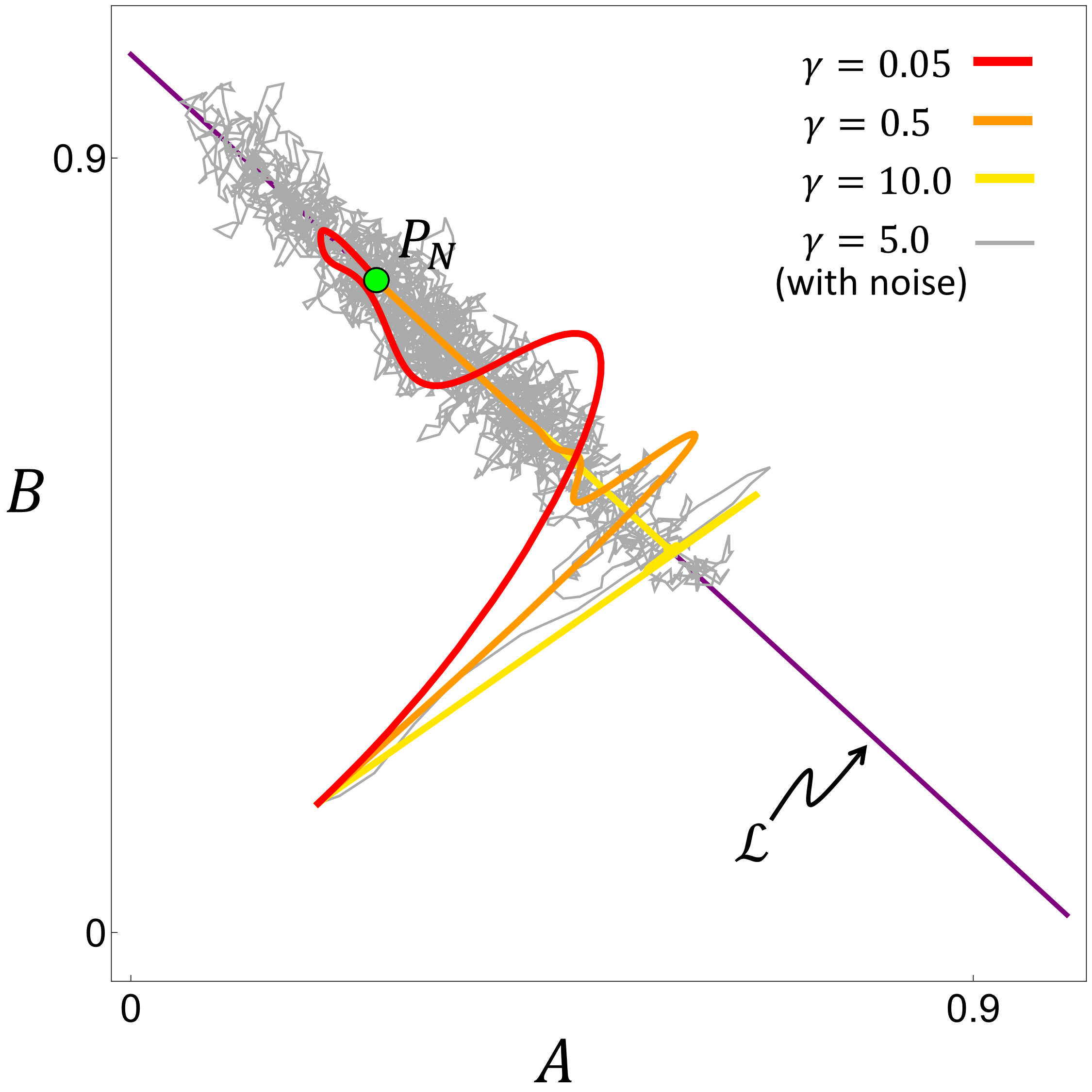}
\caption{\small Trajectories of the system projected in the $A$-$B$ plane, with parameters inside the coexistence region (by holding the position of $P_N$ fixed). As $\gamma$ increases, the system tends to approach the line $\mathcal{L}$ quickly and then crawl along it. The grey trajectory is with independent Gaussian white noise ($\sim \mathcal{N}(0,0.5)$) added to each variable's dynamics. Noise causes the system to diffuse along $\mathcal{L}$ for a long transient period before coming to the equilibrium point $P_N$. Parameters used here are $(a_0, a_1, a_2, b_0, b_1, b_2) = (0.2, 0.8, 0.5, 0.2, 0.6, 0.9)$.}
\label{fig:timenoise}
\end{figure}

An interesting behavior is seen when we examine a point inside the shrinking coexistence region as $\gamma$ is increased. Typical trajectories of the system for such parameter values are shown in Fig.~\ref{fig:timenoise}. As $\gamma$ increases, the system relaxes to the line $\mathcal{L}$ quickly, then slowly crawls along it towards the final equilibrium point $P_N$. This is because increasing $\gamma$ increases the speed that $\lambda$ relaxes to $\lambda^*$, and when $\lambda \to \lambda^*$, $\mathcal{L}$ becomes marginally stable. Therefore, the attraction to $\mathcal{L}$ in the perpendicular direction is strong, but the attraction towards the equilibrium point along $\mathcal{L}$ is weak. This leads to a long transient behavior that makes the system appear to reach no equilibrium in a limited time \cite{hastings:2018,morozov:2020}. It is especially true when there is noise in the dynamics, which will cause the system to diffuse along $\mathcal{L}$ with only a weak drift towards the final equilibrium (Fig.~\ref{fig:timenoise}). Thus, the introduction of a fast timescale (quick relaxation of $\lambda$ due to a large $\gamma$) actually results in a long transient.

\subsection{Adaptive heterogeneity} \label{sec:adaptive}

A third kind of heterogeneity we consider is the change of interactions in time. By this we mean an individual can actively change its interaction strength with others in response to certain conditions. This kind of response is often invoked in models of adaptive foraging behavior, where individuals choose appropriate actions to maximize some form of fitness \cite{abrams:1992,abrams:2010}. For example, we may consider two behaviors, resting and foraging, as our prey types. Different predators may prefer to strike when the prey is doing different things. In response, the prey may choose to do one thing or the other depending on the current abundances of different predators. Such behavioral modulation is seen, for example, in systems of predatory spiders and grasshoppers \cite{beckerman:1997}. Phenotypic plasticity is also seen in plant tissues in response to consumers \cite{agrawal:1998,lee:2011,lee:2010}.

This kind of ``adaptive'' heterogeneity can be modeled by having switching rates $\eta_1$ and $\eta_2$ that are time-dependent. Let us assume that the prey species tries to maximize its population growth rate by switching to the more favorable type. From Eq.~(\ref{eq:C-dot}), we see that the growth rate of $C$ depends linearly on the composition $\lambda$ with a coefficient $u(A,B) \equiv (a_1 - a_2) A + (b_1 - b_2) B$. Therefore, when this coefficient is positive, it is favorable for $C$ to increase $\lambda$ by switching to type $C_2$. This can be achieved by having a positive switching rate $\eta_2$ whenever $u(A,B) > 0$. Similarly, whenever $u(A,B) < 0$, it is favorable for $C$ to switch to type $C_1$ by having a positive $\eta_1$. In this way, the heterogeneity of the prey population will be constantly adapting to the predator densities. We model such adaptive switching by making $\eta_1$ and $\eta_2$ functions of the coefficient $u(A,B)$, e.g., $\eta_2(u) = 1/(1+\mathrm{e}^{-\kappa u})$ and $\eta_1 (u) = 1 - \eta_2(u)$. The sigmoidal form of the functions means that the switching rate in the favorable direction for $C$ is turned on quickly, while the other direction is turned off. The parameter $\kappa$ controls the sharpness of this transition.

\begin{figure}
\centering
\includegraphics[width=0.5\textwidth]{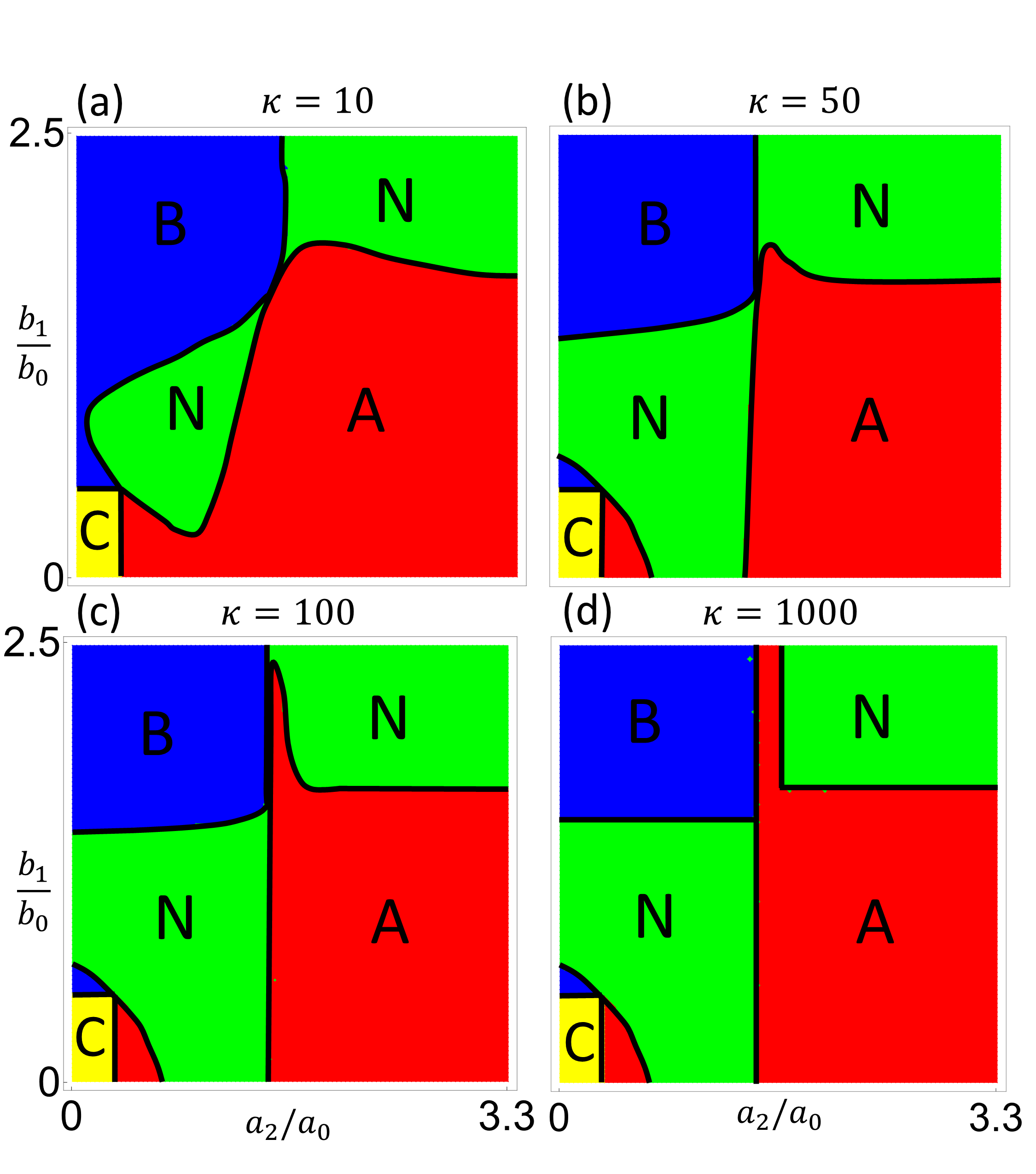}
\caption{\small Phase diagrams showing how the coexistence region changes when the switching rates $\eta_i$ dynamically adapt to the predator densities, so as to maximize the growth rate of $C$. (a--d) As the sharpness $\kappa$ of the sigmoidal decision function is increased, the region of coexistence expands. In the limit $\kappa \to \infty$, the condition for coexistence becomes quite simple (see Appendix). Parameters used here are $(a_0, a_1, b_0, b_2) = (0.3, 0.5, 0.4, 0.6)$.  }
\label{fig:eta_dot}
\end{figure}

Phase diagrams for the system with different values of $\kappa$ are shown in Figure~\ref{fig:eta_dot}. A larger $\kappa$ means the prey adapts its composition faster and more optimally, which causes the coexistence region to expand. In the extreme limit, the system changes its dynamics instantaneously whenever it crosses the boundary where $u(A,B) = 0$, like in a hybrid system \cite{goebel:2009}. Such a system can still reach a stable equilibrium that lies on the boundary, if the flow on each side of the boundary points towards the other side \cite{filippov:1988}. This is what happens in our system and, interestingly, the equilibrium is the same three-species coexistence point $P_N$ as in the previous scenarios. The region of coexistence turns out to be largest in this limit (Figure~\ref{fig:eta_dot}d).

Our results again demonstrate that the coexistence of the predators can be viewed as a by-product of the prey's strategy to maximize its own benefit. The time-dependent case studied here represents a strategy that involves the prey evaluating the risk posed by different predators. This is in contrast to the scenarios studied above, where the prey population passively creates phenotypic heterogeneity regardless of the presence of the predators. These two types of behavior are analogous to the two strategies studied for adaptation in varying environments, i.e., sensing and bet-hedging \cite{Kussell2005a, Donaldson-Matasci2013}. The former requires accessing information about the current environment to make optimal decisions, whereas the latter relies on maintaining a diverse population to reduce detrimental effects caused by environmental changes. Here the varying abundances of the predators play a similar role as the varying environment. From this point of view, the heterogeneous interactions studied here can be a strategy of the prey species that is evolutionarily favorable.

\section{Discussion} \label{sec:discussion}

Exploitative competition is a basic motif in the modeling of trophic interactions. However, the exclusion of all but one consumer that share a common resource as implied by such models is at odds with the high diversity of species seen in natural ecosystems. This is known as the ``paradox of the plankton'' \cite{hutchinson:1961}, for which many explanations have been considered \cite{chessen:2000, Barabas2016a, Chesson2018, levine:2017, saavedra:2017}. Our results suggest a new mechanism for species coexistence in these ecosystems through heterogeneous interactions between the predators and prey. Our model is general enough to describe many types of trait differentiation within a species, including phenotypic polymorphism and switching, spatial localization and migration, as well as behavioral changes and foraging strategies. Our results are related to the following ecological concepts that have been studied previously.

\subsection{Resources competition and nullcline analysis} \label{sec:nullcline}

\begin{figure}
\centering
\includegraphics[width=0.5\textwidth]{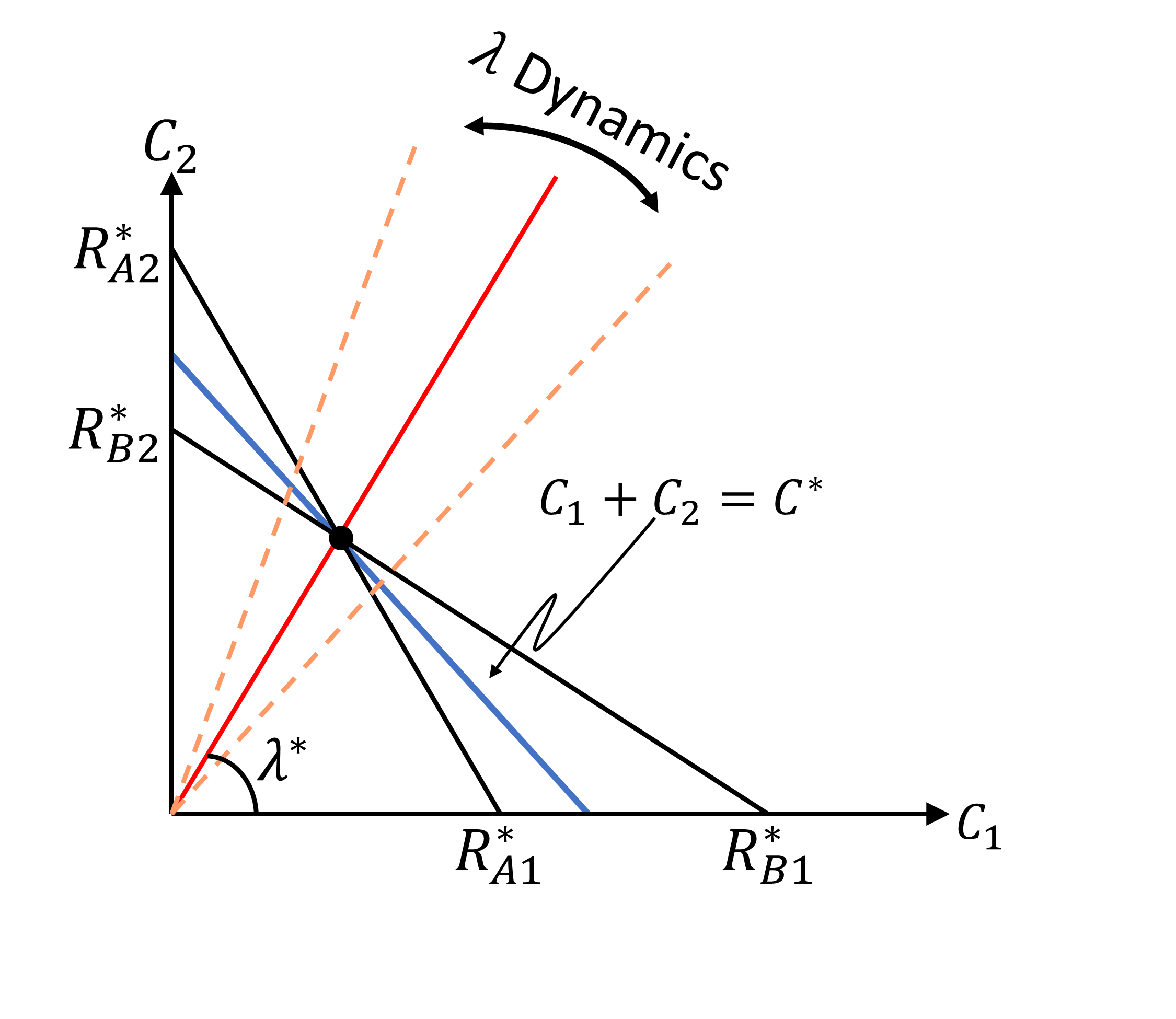}
\caption{\small A schematic for nullcline analysis showing the prey subpopulations as two resources. Nullclines for $A$ and $B$ are shown in black. Total population of $C$ is represented by a line with slope $-1$. Prey composition $\lambda$ is represented by the angle of a line going through the origin. The equilibrium values of $C^*$ and $\lambda^*$ at three-species coexistence is determined by the intersection of the nullclines. However, for our models of heterogeneous interactions, the dynamics of $\lambda$ and its stability at $\lambda^*$ are undetermined by this picture.}
\label{fig:rstar}
\end{figure}

If we think of the two subpopulations of the prey as two resources, then the competition between the two predators (consumers) can be analyzed using nullclines. A nullcline is a contour line in the space of both resource densities ($C_1$ and $C_2$), along which the net growth rate of a consumer ($\dot{A}$ or $\dot{B}$) is zero. As illustrated in Figure~\ref{fig:rstar}, there is one nullcline for each consumer ($A$ and $B$). The nullcline is a generalization of the $R^*$ value for a single (homogeneous) resource, and represents the minimal \textit{combination} of resource levels required for a consumer to sustain its population. For resources that vary independently, the persistence of the consumers can be determined from this picture. For example, if the resource levels are above the $A$ nullcline and below the $B$ nullcline, then the $A$ population can grow but the $B$ population will decline, leading to the exclusion of $B$ by $A$. By such analysis, the two consumers can coexist only if the resource levels are precisely at the intersection of the nullclines \cite{tilman:1982}.

The nullcline analysis helps determine the location of the coexistent equilibrium in the parameter space. However, the stability of this equilibrium point cannot be derived from the picture, because in our model the two resource types interact with each other dynamically. Indeed, they are subpopulations of the same prey species, exchanging fluxes of individuals and sharing a common carrying capacity. The ratio of the prey types ($\lambda$) is determined by the angle of lines going through the origin (Figure~\ref{fig:rstar}). The dynamics of $\lambda$ and of the total prey population $C$ both depend on the densities of $A$ and $B$ (Eqs.~(\ref{eq:C-dot},\ref{eq:lambda-dot})) and are not captured in the picture here. Nevertheless, the nullcline analysis does show that the equilibrium values of $C^*$ and $\lambda^*$ in the coexistence phase do not depend on the parameters such as $\rho$ and $\gamma$, and in fact do not depend on the form of heterogeneity at all (see Appendix). Beyond that, the main point of our model is to demonstrate that the coexistent equilibrium is in fact stable, and is robust to the different kinds of heterogeneity that we studied.

\subsection{Emergent fitness equalization vs. niche separation} \label{sec:equalization}

It is important to note that the coexistence of the predators shown here does not result from the separation of niches, but rather from the convergence of the predators to a common fitness, similar to the situation studied in \cite{velzen:2020}. The separation of niches would mean that each predator consumes only one prey type. Here the niches of the predators overlap because they consume both prey types but with different preferences. In \cite{velzen:2020} and in the models that we study here, an internal variable of the prey population (the phenotype composition in our case, and the defense level in \cite{velzen:2020}) is adjusted to ensure equal fitness of the predators. Here fitness is used to mean the $R^*$ values of the predators for a given internal trait value. The convergence of fitness is emergent in that it is not the result of fine-tuning the predators' consumption preferences (i.e., interaction strengths), but rather a result of the dynamics of the prey composition that automatically stabilizes at the critical value where $R^*$'s are equal. Thus, the coexistence of predators observed in our model is an emergent consequence of the prey having heterogeneous interaction strengths with the predators.

\subsection{Emergent facilitation and trait-mediated indirect effects} \label{sec:facilitation}

\begin{figure}
\centering
\includegraphics[width=0.5\textwidth]{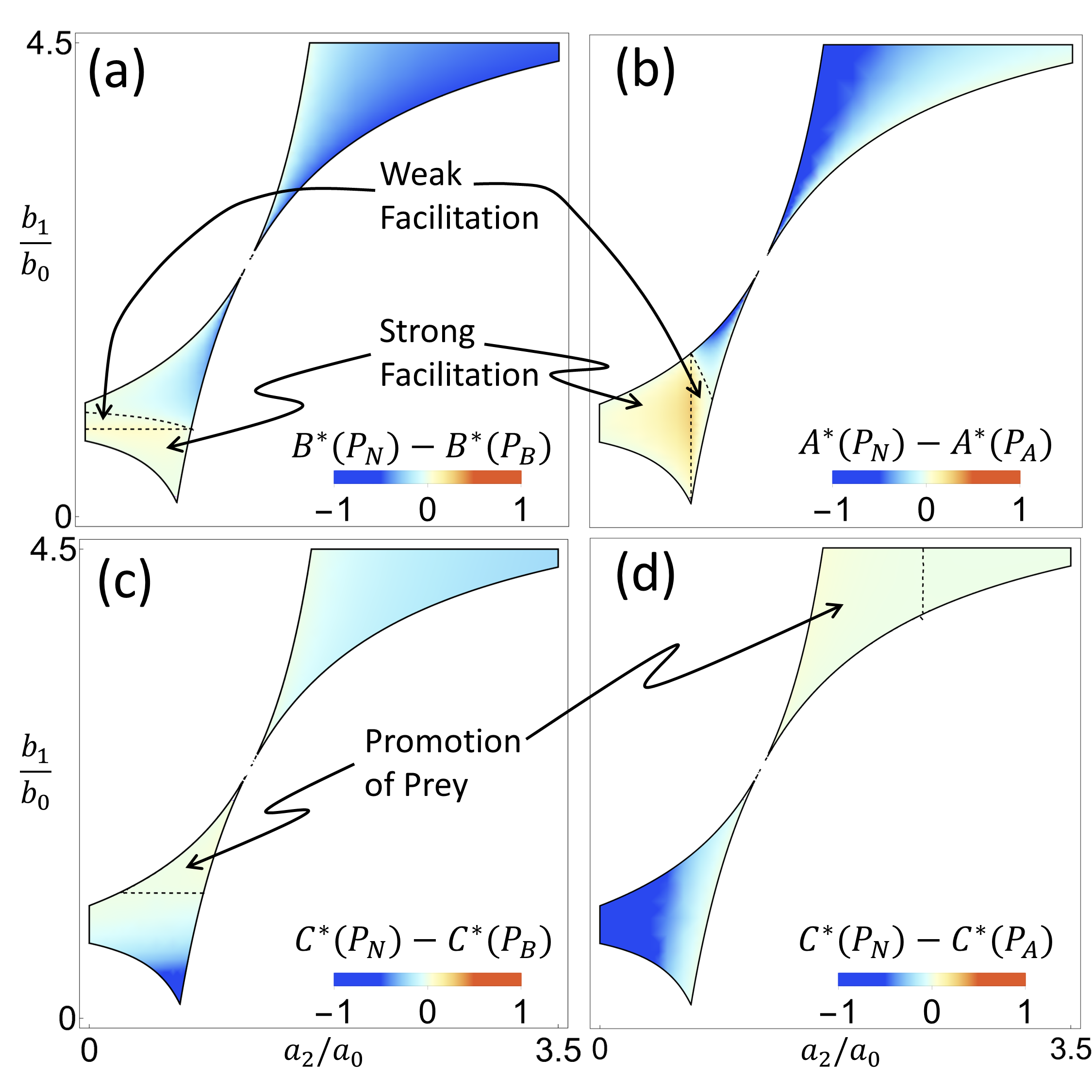}
\caption{\small Phase diagrams showing facilitation of predators by their competitor, and promotion of prey by predators. (a,b) Parameter regions of strong and weak facilitation, with (a) $B$ facilitating $A$ and (b) $A$ facilitating $B$. Color represents the difference between population levels of each predator in the presence and absence of the other. (c,d) Parameter regions showing promotion of the prey by the addition of a second predator, with (c) $B$ promoting $C$ and (d) $A$ promoting $C$. Color represents the difference in population levels of the prey $C$ in the presence of both or only one predator. Parameter values used here are $(a_0, a_1, b_0, b_2, \rho) = (0.25, 0.6, 0.5, 0.6, 0.55)$.}
\label{fig:facil}
\end{figure}

Our model also demonstrates an interesting effect that one predator can have on the other, namely facilitation \cite{deroos:2008,deroos:2020}. It happens when one predator is better off in the presence of another even though they compete for the same resources. We can distinguish between two levels of facilitation: Strong facilitation is such that one predator would go extinct if the other predator is absent but can survive when both are present. Weak facilitation is such that the abundance of one predator is increased when the other predator is present rather than absent. Strong facilitation was studied in \cite{deroos:2008} in the case of adult and juvenile prey with body size differentiation. In that system, facilitation emerges as a result of two predators specializing in different prey stages and the prey stages competing with each other. Our system shares the feature that two predators prefer different subpopulations of the same prey. We found that our models exhibit both strong and weak facilitation, as show in Fig.~\ref{fig:facil}(a,b) for the case of inherent heterogeneity. Intuitively, the consumption of one prey type by one predator allows the other prey type to flourish, thus helping the other predator.

We can interpret such emergent facilitation as a trait-mediated indirect effect \cite{wootton:1994,holt:2012}. The interaction between the predator $A$ and the prey $C$ causes a shift in the composition of $C$, which can be considered a trait of $C$ on the population level. This trait change then affects the overall interaction between $C$ and the other predator $B$. Therefore, in a coarse-grained picture where the subpopulations of $C$ are lumped together (Figure~\ref{fig:variables}c), it is as if one species $A$ modifies the effective interaction strength $b_\textrm{eff}(\lambda)$ between the other two species $B$ and $C$, and similarly for $B$ that modifies the interaction between $A$ and $C$. Here the interaction modification by one predator can be either detrimental to the other predator, or beneficial as in the cases of facilitation.

\subsection{Multiple-predator effects and emergent promotion of prey}
\label{sec:promotion}

In addition to the facilitation between predators shown above, our model also exhibits a surprising ``promotion'' of the prey population by a predator. By this we mean that the equilibrium population of the prey is higher in the presence of both predators than in the presence of just one. This will be an extreme case of subadditive effects on the prey from multiple predators \cite{sih:1999, mccoy:2012}, to the extent that the total effect from two predators is lower than that from one predator alone. Figure~\ref{fig:facil}(c,d) show the regions in parameter space where there is promotion of the prey $C$ by predator $A$ and $B$ respectively. This surprising effect arises when both predators prefer to consume the same prey type, and the promoting predator has a stronger preference than the other. Heuristically, if we start with only the latter predator in equilibrium with the prey, then adding the promoting predator would push the composition of the prey away from the preferred type. This compositional shift then inhibits both predators and allows the prey to reach a larger population at the new equilibrium. Mathematical conditions for such emergent promotion to happen are given in the Appendix.

\section{Conclusion} \label{sec:conclusion}

Heterogeneity is natural in biological systems, as each individual organism possesses a large number of traits that are influenced one way or the other by developmental noise or environmental variation. Every individual interacts with other individuals and the environment somewhat differently due to its unique combination of traits. The heterogeneity in such interactions are generally treated in one of two ways. It is ignored in some cases where the system is treated as being well-mixed or homogeneous, such as in the classic Lotka-Volterra model \cite{lotka:1910,lotka:1925}. In other cases the system is treated as ``disordered'' such that the interaction strengths are randomly drawn from some probability distribution \cite{may:1972,gravel:2016,barbier:2021}. Here, we have addressed the heterogeneity of biological interactions in a more structured framework that considers subpopulations of a species with a dynamic composition. We have demonstrated nontrivial consequences of having heterogeneous interactions. We expect such ecosystems to exhibit stronger persistence and richer diversity. We find that one predator can facilitate another both qualitatively and quantitatively through trait-mediated indirect effects. And we show that a prey species can change its population composition to benefit its own growth under multiple predators. These effects are apparent in our model with only one species of a small ecosystem exhibiting a single differentiating trait. In real natural ecosystems, there is a massively higher dimensionality in both the number of traits within a species that can vary among individuals and the number of species that can exhibit such trait differentiation. We expect such high dimensionality to lead to more manifestations of the effects illustrated here.

We have focused here on biological and ecological systems, but heterogeneous interactions can have nontrivial effects on dynamical systems in general. Here we have seen new attractors appear where they could not in a homogeneous system. We have seen new timescales introduced to the system that can result in slow convergence to the equilibrium. We have even seen the transformation of the dynamical system from a simple differential system to one which exhibits flow switching depending on the state of the system \cite{lou:2008}. It is less common for traditional systems in physics to have heterogeneous interaction constants. However, non-uniform parameters can be important in systems involving a large number of mesoscopic components. In such systems, the parameters of the components are often non-identical, and the shape or width of the parameter distribution can have strong effects on the behavior of the system. This can be seen in systems of coupled oscillators \cite{strogatz:2000, ulrichs:2009, jackson:2021} where the spread of the natural frequencies of the oscillators can lead to different degrees of synchronization. The effect of inhomogeneous parameters can also be seen in systems of colloids \cite{guram:2017}, where differences in size and mass among large numbers of colloids can lead to intermittent behavior with periods of ordered lattice configuration and more random motion. These are examples of systems where differences among interacting units of a system can have qualitative effects on the overall system behavior. As we have shown, incorporating such differences between the units can be important for predicting the outcome of these complex dynamics.

\appendix
\setcounter{figure}{0}
\renewcommand\thefigure{A\arabic{figure}}
\setcounter{equation}{0}
\renewcommand\theequation{A\arabic{equation}}

\section{Methods}

Our model of two predators and one prey that differentiates into two types is described by Eqs.~(\ref{eq:A-dot}--\ref{eq:lambda-dot}) in the main text. Changing variables from $C_1$ and $C_2$ to $C \equiv C_1 + C_2$ and $\lambda \equiv C_2 / (C_1 + C_2)$ leads to the equations:
\begin{subequations}
\begin{align}
\dot{A} &= A \big( \big( \alpha_{A1} (1-\lambda) + \alpha_{A2} \lambda \big) C - \beta_A \big) \\
\dot{B} &= B \big( \big( \alpha_{B1} (1-\lambda) + \alpha_{B2} \lambda \big) C - \beta_B \big) \\
\dot{C} &= C \big( \beta_C-\alpha_{CC}C - \varepsilon_A \big( \alpha_{A1} (1-\lambda) + \alpha_{A2} \lambda \big) A - \varepsilon_B \big( \alpha_{B1} (1-\lambda) + \alpha_{B2} \lambda \big) B \big) \\
\dot{\lambda} &= \lambda (1-\lambda) \big( \varepsilon_A (\alpha_{A1} - \alpha_{A2}) A + \varepsilon_B (\alpha_{B1} - \alpha_{B2}) B \big) + (1-\lambda) \sigma_1 - \lambda \sigma_2
\end{align}
\end{subequations}
To simplify, we then rescale the variables by:
\begin{align*}
t \leftarrow \beta_C t, \; A \leftarrow \frac{\varepsilon_A \alpha_{CC}}{\beta_C} A, \; B \leftarrow \frac{\varepsilon_B \alpha_{CC}}{\beta_C} B, \; C \leftarrow \frac{\alpha_{CC}}{\beta_C} C
\end{align*}
and rename the parameters as:
\begin{align*}
a_0 = \frac{\beta_A}{\beta_C}, \; a_1 = \frac{\alpha_{A C_1}}{\alpha_{CC}}, \; a_2 = \frac{\alpha_{A C_2}}{\alpha_{CC}}, \; b_0 = \frac{\beta_B}{\beta_C}, \; b_1 = \frac{\alpha_{B C_1}}{\alpha_{CC}}, \; b_2 = \frac{\alpha_{A C_2}}{\alpha_{CC}}, \; \eta_1 = \frac{\sigma_1}{\beta_{C}}, \; \eta_2 = \frac{\sigma_2}{\beta_{C}}
\end{align*}
After these transformations, we arrive at Eqs.~(\ref{eq:A-dot}-\ref{eq:lambda-dot}) in the main text.

In the case of inherent heterogeneity, the dynamical equations for $C_1$ and $C_2$ are:
\begin{subequations}
\begin{align}
\dot{C_1} &= (1-\rho) \beta_C C - \alpha_{CC} C_1 C - \varepsilon_A \alpha_{A1} C_1 A- \varepsilon_B \alpha_{B1} C_1 B  \\
\dot{C_2} &= \rho \beta_C C - \alpha_{CC} C_2 C - \varepsilon_A \alpha_{A2} C_2 A - \varepsilon_B \alpha_{B2} C_2 B
\end{align}
\end{subequations}
Using the same transformations and rescaling as above, these equations become the same as Eqs.~(\ref{eq:C-dot}-\ref{eq:lambda-dot}) in the main text with $\eta_1 = \rho (1-C)$ and $\eta_2 = (1-\rho)(1-C)$.

The locations of the equilibrium points in the $A$-$B$-$C$ space are:
\begin{align*}
P_O &= \bigg( 0, 0, 0, \frac{\eta_1}{\eta_1+\eta_2} \bigg) \\
P_C &= \bigg( 0, 0, 1, \frac{\eta_1}{\eta_1+\eta_2} \bigg) \\
P_A &= \bigg( \frac{a_1 (1-\lambda) + a_2 \lambda - a_0}{(a_1 (1-\lambda) + a_2 \lambda)^2}, 0, \frac{a_0}{a_1 (1-\lambda) + a_2 \lambda}, \lambda \bigg)
\end{align*}
where $\lambda$ is the solution to the cubic equation
\begin{align*}
\lambda (1-\lambda) (a_1 - a_2) (a_1 (1-\lambda) + a_2 \lambda - a_0) + (\eta_1 (1-\lambda) - \eta_2 \lambda) (a_1 (1-\lambda) + a_2 \lambda)^2 = 0
\end{align*}
And $P_B$ has the same expression except with parameters associated with species $B$. The location of $P_N$ depends on the form of $\eta_i$'s, but it always lies on the line $\mathcal{L}$ given by $\{ a_0 A + b_0 B = C^* (1 - C^*), C = C^* \}$. For fixed $\eta_i$'s,
\begin{align*}
P_N &= \Bigg( \frac{1}{a_0} \bigg( \frac{\tilde{b}_2 - \tilde{b}_1}{\tilde{a}_1 - \tilde{a}_2 + \tilde{b}_2 - \tilde{b}_1} C^* (1 - C^*) - \bigg( \frac{\eta_1}{\tilde{a}_1 - \tilde{b}_1} - \frac{\eta_2}{\tilde{b}_2 - \tilde{a}_2} \bigg) \bigg), \\
& \frac{1}{b_0} \bigg(\frac{\tilde{a}_1 - \tilde{a}_2}{\tilde{a}_1 - \tilde{a}_2 + \tilde{b}_2 - \tilde{b}_1} C^* (1 - C^*) + \bigg( \frac{\eta_1}{\tilde{a}_1 - \tilde{b}_1} - \frac{\eta_2}{\tilde{b}_2 - \tilde{a}_2} \bigg) \bigg), C^*, \lambda^* \Bigg)
\end{align*}
where
\begin{align*}
C^* &= \frac{\tilde{a}_1 - \tilde{a}_2 + \tilde{b}_2 - \tilde{b}_1}{\tilde{a}_1 \tilde{b}_2 - \tilde{a}_2 \tilde{b}_1} \\
\lambda^* &= \frac{\tilde{a}_1 - \tilde{b}_1}{\tilde{a}_1 - \tilde{a}_2 + \tilde{b}_2 - \tilde{b}_1}
\end{align*}
with $\tilde{a}_1=a_1/a_0$, $\tilde{a}_2=a_2/a_0$, $\tilde{b}_1=b_1/b_0$, and $\tilde{b}_2=b_2/a_0$. The solutions for the inherent case are recovered by making the substitutions for $\eta_1$ and $\eta_2$ above, and for the adaptive case by removing the $\eta$ terms.

The stability of these equilibrium points is determined by the following rules, with the physical region $\mathcal{W} \equiv \{ A \geq 0, B \geq 0, 0 \leq C \leq 1, 0 \leq \lambda \leq 1 \}$:
\begin{itemize}
\setlength{\itemsep}{0pt}
\setlength{\itemindent}{-5pt}
\setlength{\parsep}{0pt}
\setlength{\parskip}{0pt}
\item If only the points $P_O$ and $P_C$ are in $\mathcal{W}$, then $P_C$ is stable.
\item If only $P_O$, $P_C$ and $P_A$ (or $P_B$) are in $\mathcal{W}$, then $P_A$ (or $P_B$) is stable. 
\item If $P_O$, $P_C$, $P_A$, $P_B$ are all in $\mathcal{W}$ but not $P_N$, then one of $P_A$ and $P_B$ with a smaller $C^*$ is stable.
\item If $P_N$ is in $\mathcal{W}$ then it is stable, or there is a small limit cycle enclosing it that is stable.
\end{itemize}

For completeness we list here the exact conditions for the regions of coexistence between the predators. The expression is simplest for the case of instant ($\kappa \to \infty$) adaptive heterogeneity:
\begin{align*}
\big( (\tilde{a}_2 < \tilde{b}_2) \land (\tilde{b}_1 < \tilde{b}_2) \land (C^*(P_N) < 1) \big) \lor \big( (\tilde{a}_2 > \tilde{a}_1) \land (\tilde{b}_1 > \tilde{a}_1) \big)
\end{align*}
For the other kinds of heterogeneity, the condition for coexistence is given by
\begin{align*}
(A^*(P_N) > 0) \land (B^*(P_N) > 0) \land (1 > C^*(P_N) > 0) \land (1 > \lambda^*(P_N) > 0)
\end{align*}

In the case of inherent heterogeneity, the requirement for strong facilitation of predator $B$ by predator $A$ is that $P_C$ is stable in the system when $A=0$ and $P_N$ is stable when $A$ is present. For weak facilitation it is required that $P_B$ and $P_N$ are stable, respectively, and $B^*(P_N) > B^*(P_B)$. In order to see emergent promotion, in addition to $P_N$ being stable, we need $a_1/a_0>b_1/b_0>b_2/b_0>a_2/a_0$ for $A$ to promote $C$, and $b_1/b_0>a_1/a_0>a_2/a_0>b_2/b_0$ for $B$ to promote $C$.

\bibliographystyle{unsrt}
\bibliography{jackson_arxiv_version}

\end{document}